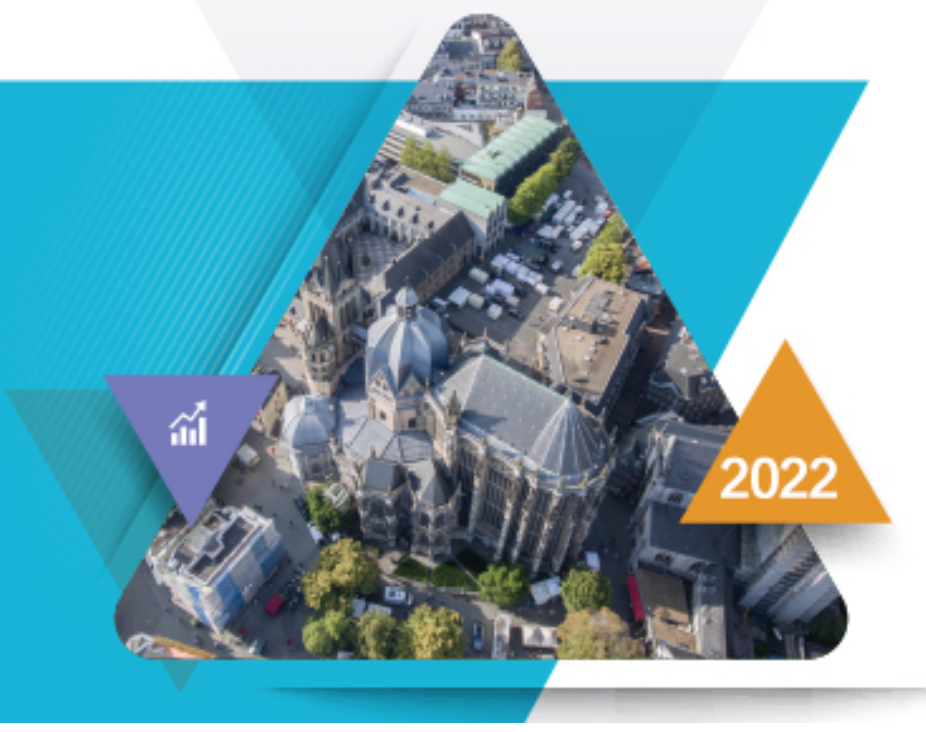

# Design and implementation of grid-connected photovoltaic power plant with the highest technical Efficiency

**Mehran Hosseinzadeh Dizaj**

Ghods Niroo Engineerong Co

PhD of Electronic Engineering, m.hoseinzadeh@ghods-niroo.com

## Abstract

Energy is a necessity and the basis of human life. With the increase in the need for energy supply in recent years, the use of fossil fuels has intensified. Environment is a basic principle for human beings. In our dear country Iran, with an average sundial of approximately 4.5 hours per day, there is the highest potential for the use of photovoltaic systems. Connected to the national power grid as a micro grid, in accordance with the standardization and the highest efficiency with PVsyst software, the system losses were identified and the necessary solutions were provided to solve them. In case of using 1 to 5 kW systems, which due to small It can be installed even on the roofs of houses. In hot cities, where electricity consumption increases in the hot seasons of the year, the pressure is removed from the national electricity distribution network, and a significant amount of greenhouse gas emissions to the environment is prevented. , At the end by plotting the current value of the project it was also proven to be economical.

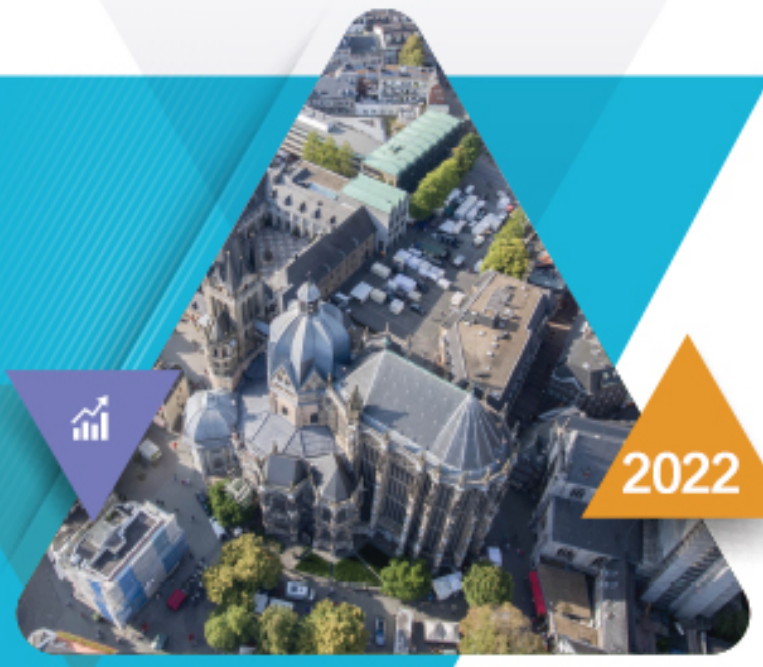

## Introduction

Today, political, economic crises, and issues such as limited fossil fuel sustainability, environmental concerns, acid rain, global warming, overcrowding, economic growth, and consumption rates are all universal topics that are widely considered by thinkers. Appropriate solutions to the proper solution of energy problems in the world, especially environmental crises, have been busy. The access of developing countries to new types of energy sources is essential for their economic development, and new research has shown that there is a direct relationship between the level of development of a country and its energy consumption. Given the limited reserves of fossil energy and the increasing level of energy consumption in the current world, it is no longer possible to rely on existing energy sources. According to current human technology, nuclear energy and hydropower are two alternative energy sources for fossil fuels. Most countries in the world realize the importance and role of various energy sources, especially renewable energy (new) in meeting current and future needs and widely, in the development of exploitation of these essential resources, extensive research and principled investments. Paying attention to the increasing need for energy consumption and reducing fossil energy sources, the need to protect the environment, reduce air pollution, electricity supply restrictions and fuel supply for remote areas and villages, use of new energy such as wind, solar, hydrogen, biomass, Geothermal can have a special place. Considering such basic and increasing trends in the use of renewable energy and related technologies in industrial and developing countries in Iran, it is necessary to develop basic strategies and programs.

## Photovoltaic solar systems:

In recent years, the use of new energy sources has grown significantly due to the limitations and rising prices of fossil fuels as well as environmental issues. Solar energy is the largest source of renewable energy on Earth, available directly and indirectly. There are two ways to benefit from solar energy:

1. Using sunlight and converting it into electricity through photovoltaic cells
2. Using the thermal energy of the sun and converting it into other types of energy or using it directly

## Photovoltaic systems consisting of three general parts:

Solar panel, converter and battery, convert solar energy into electricity without any pollution. The conversion of solar energy into electricity is done by a panel or the same as photovoltaic cells, and in order to use it in domestic applications, it converts the generated electricity into alternating current. Finally, with the help of the battery, the extra electrical energy produced in the system can be stored. Other peripherals of the system include wires, circuit breakers and support structures.

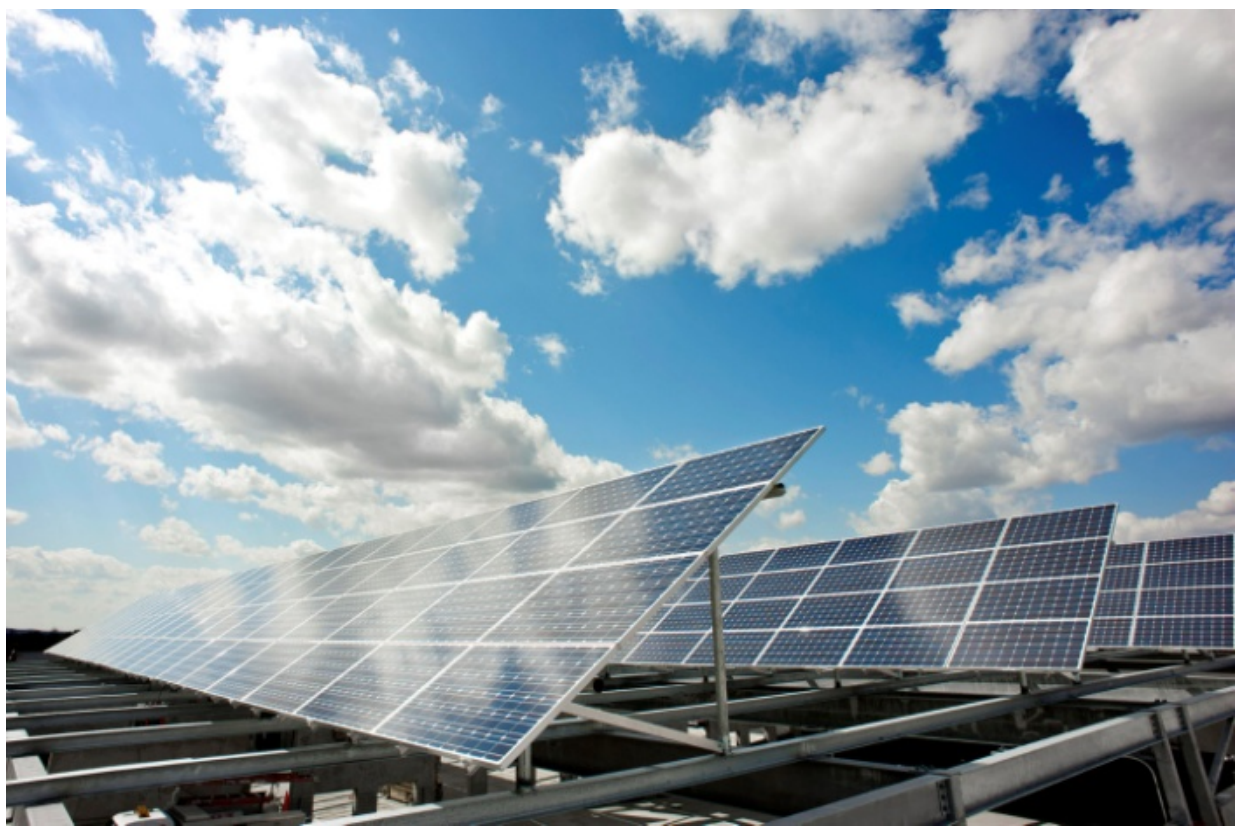

**Figure 1- Solar arrays**

## User classification of photovoltaic systems:

In general, photovoltaic systems are classified into two groups according to their application: grid-connected units and separate units from the grid. In a grid-connected system, electricity generated from solar energy will be injected into the national grid. In other words, in this system, the user supplies his generated electricity to the New Energy Organization (Ministry of Energy). Photovoltaic systems connected to the national grid in a centralized or decentralized manner are used to strengthen the national grid and prevent electrical pressure on power plants during the day. The advantages of this system include easy installation, high efficiency and no need for complex peripherals. Figure 2 shows a grid-connected photovoltaic system.

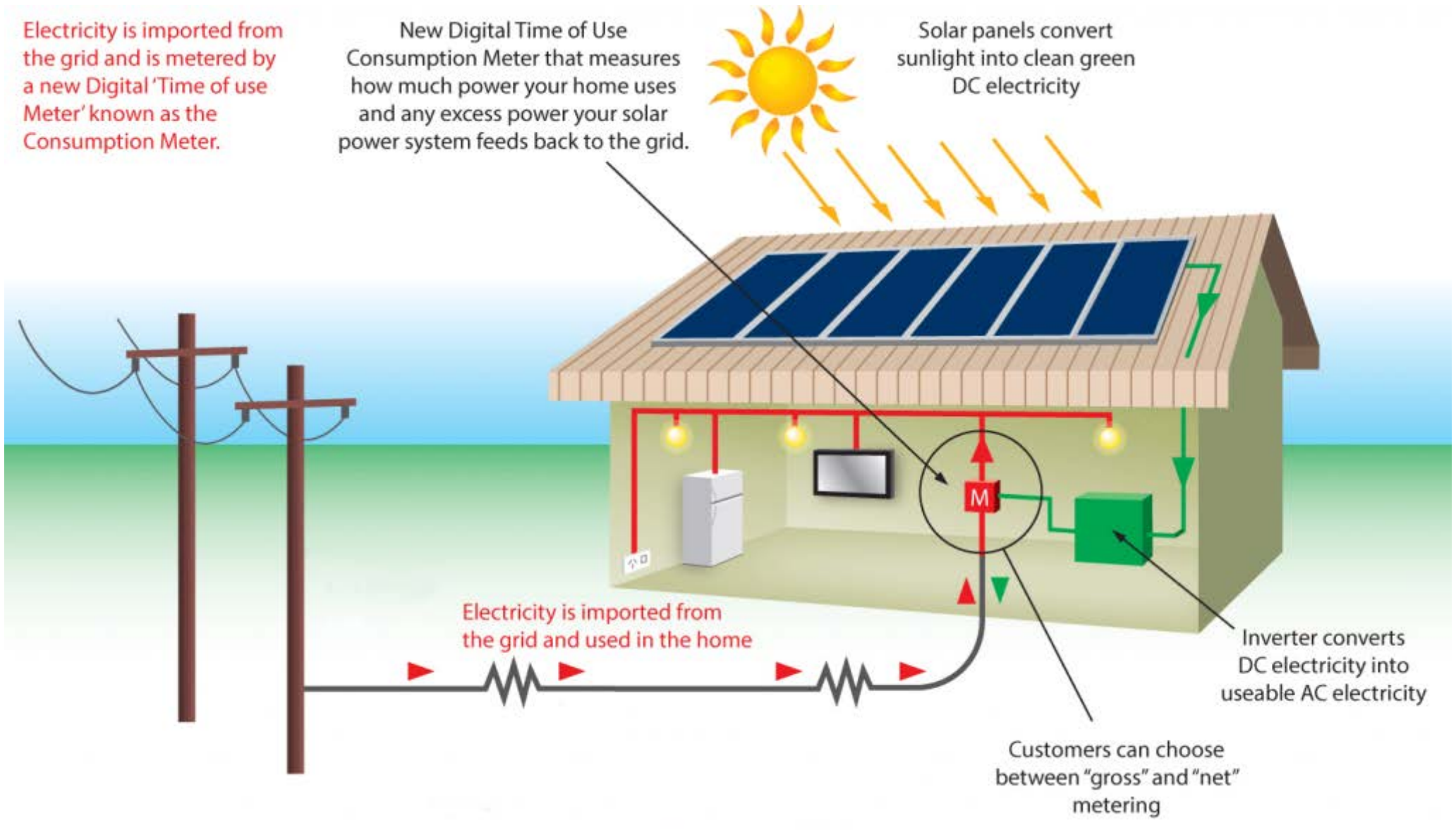

**Figure 2 - Photovoltaic system connected to the national grid**

Normally, this type of networking does not require a battery to store electrical energy, but in some cases, energy storage systems, which are mainly batteries, are used to increase the reliability of the network. Therefore, systems connected to the national grid can be classified into two groups with storage system and without storage system. The design of grid-connected photovoltaic systems is such that they generate power simultaneously and in parallel with the national grid. In this method, there is a two-way connection between the photovoltaic systems and the power grid, so that if the DC electricity produced by the photovoltaic systems exceeds the local load requirement, the excess is injected into the national grid, so at night and when Climatic conditions It is not possible to use sunlight. The electricity required by the site is provided by the national electricity network. In general, the electricity generated after being converted by the inverter for grid-connected systems and using special two-way meters, is injected into the national grid. If for any reason the grid is cut off, it is necessary that the solar unit also stops production and cut off loads. Figure 3 is a block diagram of an example of a photovoltaic system without a storage system.

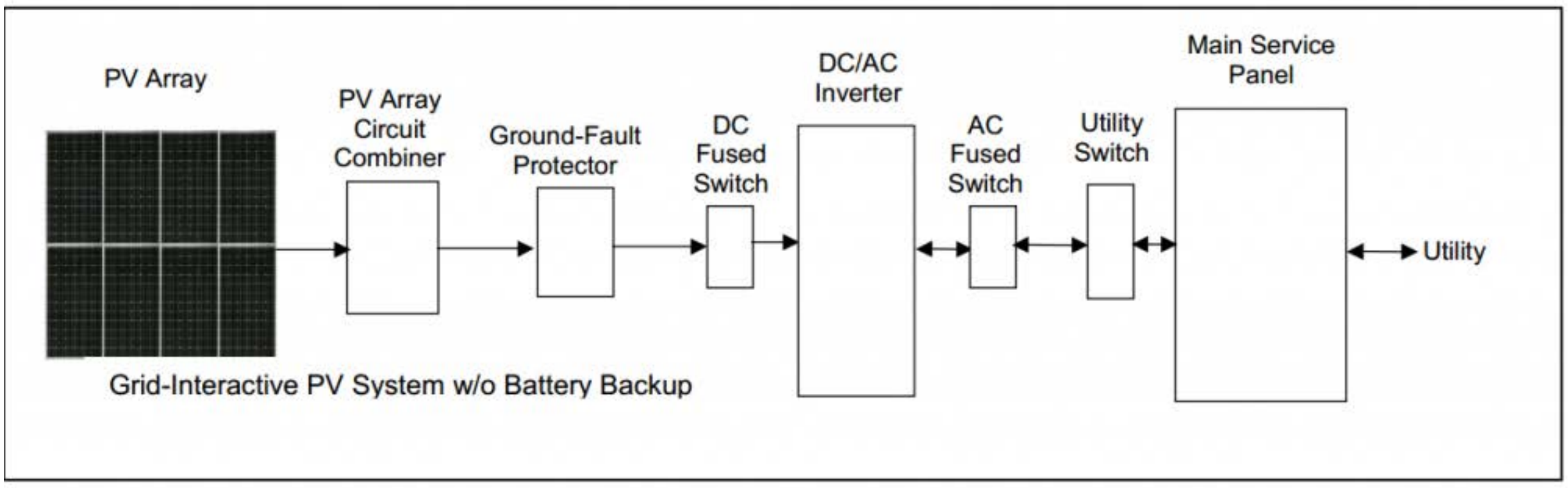

**Figure 3 - Diagram of a photovoltaic system without a battery**

In grid-connected photovoltaic systems with a storage battery, the electronic power converter is powered by a solar array whose output is connected to the battery as a local charge and sensitive load energy storage. Grid-connected photovoltaic systems are equipped with storage systems suitable for residential homes and small commercial neighborhoods because these systems use stored energy for sensitive loads such as refrigerators, lighting, elevators and the like. Figure 4 is a block diagram of this group of photovoltaic systems.

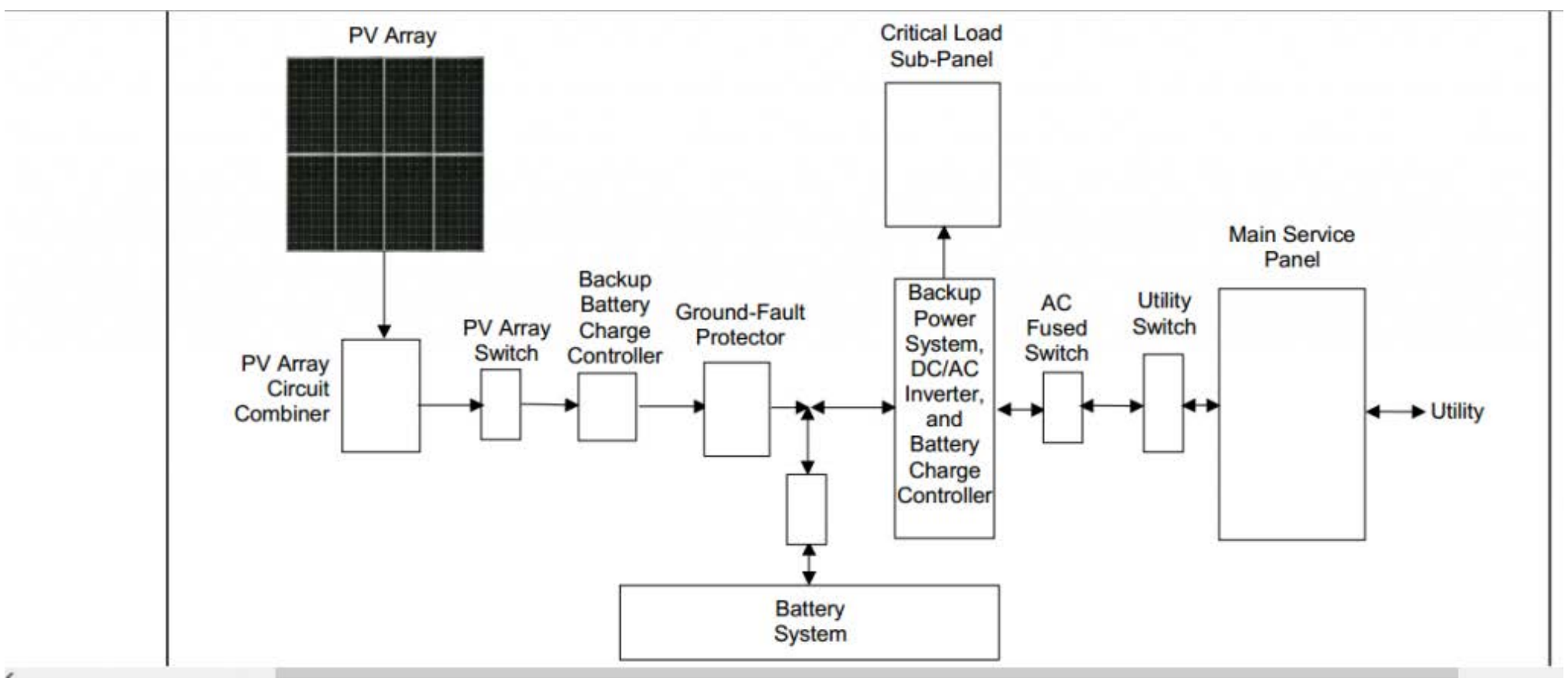

**Figure 4 - Diagram of a photovoltaic system with a battery storage system**

Grid-independent photovoltaic system is also known as island system due to its independent nature. In this type of solar system, the generated electricity is solely responsible for charging the batteries. Therefore, to design and calculate the number of photovoltaic arrays in such units, it is necessary to calculate the load model and the total power required for the load in a 24-hour period. On the other hand, in unfavorable weather conditions, it is necessary for the energy storage system to be able to feed the entire load of the system without using the energy of the photovoltaic system for several days. The main application of such units is in places where the main power network is not available Connecting to a power grid costs a lot of money. For example, in mountainous telecommunication sites, nomadic areas, rural cottages, and in general to meet the electrical needs of areas that do not have a national electricity grid, a grid-independent photovoltaic system can be used. Figure 4 shows a grid-independent photovoltaic system. The main components of this type of solar system are:

• Arrays mounted on the roof or mounted on the ground

• Battery charge controller

• Converter to supply power consumption

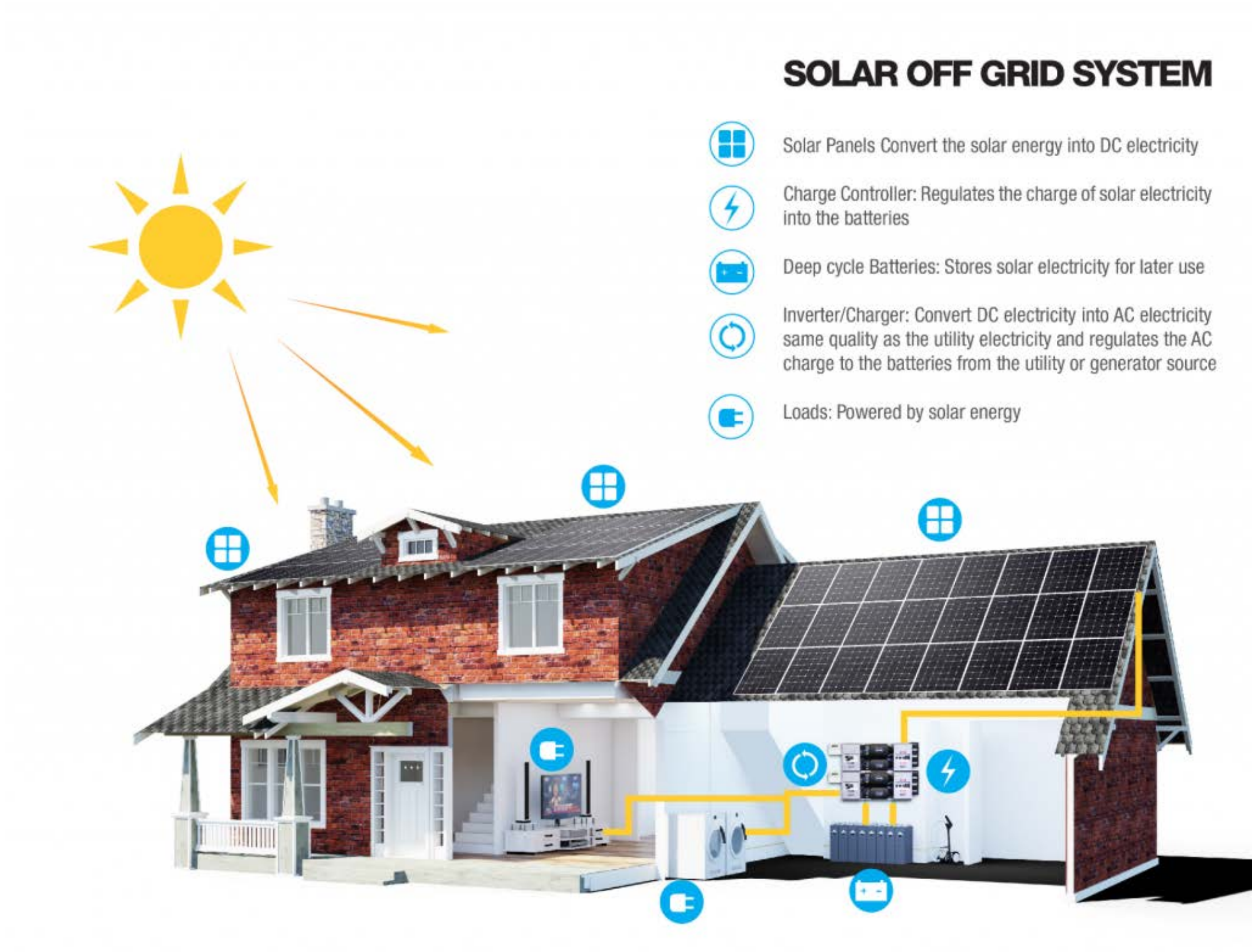

**Figure 5 - Grid-independent photovoltaic system**

**1: Solar panels:**

This part is the conversion of solar radiant energy into electrical energy without mechanical intermediation. It should be noted that the output current and voltage of these panels are DC. These panels are made to withstand all environmental hardships such as extreme polar cold, desert heat, tropical humidity and strong winds. However, these devices are made of glass and may break due to heavy impacts.

**2: Optimal power generation or control:**

This section controls all the specifications of the system and injects or controls the production capacity of the panels according to the design and the consumer's need for load or battery. It should be noted that in this section, the specifications and components according to The needs of the electric charge, the consumer as well as the local weather conditions change. Therefore, possible failure in any part or information related to each part can be taken from the control part. This set consists of several sub-sections or sections, which include: battery, charge control, MPPT, inverter and control system. It should be noted that not all the mentioned parts are necessarily used for each consumer, but according to the specifications and needs of each consumer, the desired power generation section consists of some of the mentioned sub-sections. Therefore, the tasks of the controller are as follows:

- Match the performance of all system components (including MPPT, charge control and..)
- Command to different sections when necessary
- Collect information from system performance
- Notification of system components.
- Protection of the whole system
- Earth system protection

**3: Energy saver and battery:**

The sun's radiant energy varies throughout the day, so many solar energy applications require an energy storage source to store the electrical energy produced by sunlight during the day and to use it during peak times or in the absence of the sun.

**3-1: Charging control and load control unit:**

The main task of this section is to control the charge and discharge status of batteries. In order to use their maximum useful life, it consists of two charging sections and a load voltage control unit. The charging department controls the charge status of the batteries in terms of input current and voltage, ambient temperature and electrolyte concentration, etc., and when necessary, according to the designs, performs the necessary performance in accordance with the condition and condition of the batteries on the system. Slow in order to increase the useful life and allow the consumer to use the maximum available capacity of batteries. The other part of the task is to regulate and control the discharge cycle of the batteries and prevent the reduction of the life and wear of the batteries.

## In short, the function of this device is:

1. Panel output voltage test
2. Test the output current of the panels
3. Battery output voltage test
4. Battery output current test
5. Ambient temperature test
6. Electrolyte concentration test of batteries
7. Deciding to turn off or on the voltage and output current of the panels to charge the batteries
8. Deciding to disconnect or connect the output voltage and current of the panels to the consumer

### 3-2: MPPT:

This system is a DC-DC converter that provides impedance matching between the dynamic resistance of solar panels and the consumer. This system can be used in standalone systems as well as in systems connected to the national electricity grid.

### 3-3: Inverter, DC / AC converter:

Power conversion from DC to AC is done by a converter (inverter). In photovoltaic systems, the resulting electricity is DC, and since most of the loads in industry and electric applications work with AC power, this electricity can be converted by an inverter device and its characteristics such as voltage and frequency with Consumer components matched.

### 3-4: Energy storage system:

The presence of storage source in the photovoltaic system is very important an photovoltaic systems can be divided into two general categories based on storage systems with storage source and without storage source.

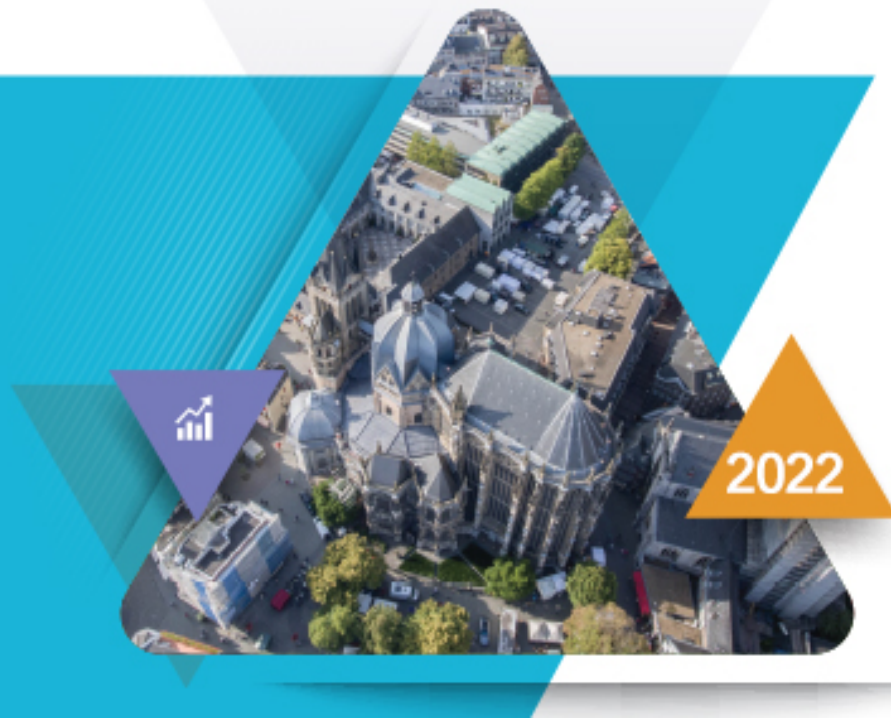

## Type of photovoltaic panel used:

High-consumption solar panels are divided into single-crystal and polycrystalline to create a photovoltaic array. Crystals are usually less than 15% and the yield of single crystals is between 15 and 19%. Arrays together, this drop affects the whole array and we will face a decrease in their efficiency. In hot areas, we need high voltage efficiency. So the selected panels for project arrays are of monocrystalline type, in installing the panels we have to use brackets (frame and bases of panels), according to the table we can choose the material of the bracket.

| Gender | Advantages | Disadvantages | Recommended use |
|---|---|---|---|
| Wood | Easy access, cheap | Low resistance to moisture | Hot and dry places, low wind intensity |
| Iron | Strong, cheap | Caries due to moisture | Moderate, warm and dry places |
| Aluminum | Easy to work with, moisture resistant | Being more expensive | For wet places |
| Steel | Moisture resistant, very | Expensive, hard to work with | For different places |

**Table 1: Bracket material and their advantages and disadvantages**

If a photovoltaic system is installed on roofs, the strength of the roof should be checked. If you are weak, be sure to strengthen to be able to weigh the required weight of solar panels, which usually have an average weight of less than 20 kg per square meter Bear. If the installation site is logically strong with strong winds, be sure to secure the brackets, it is better the panels should be installed at a short distance from each other. In order for the panels to be cooled by the surrounding air. Must be their distance from the ground, Observe so that the air circulates around them to cool them.

## Simulation and design of photovoltaic power plant:

320 watt panels, of national production type, 20 panels arranged in 2 8-digit strings. 5 kW MPPT inverter this model has a high efficiency for power transmission. Fixed angle brackets are used because of their long-term strength and lower cost. Table 2 shows the average price and the most important costs for economic calculations.

| Name | Type | Number | unit price | The price of the whole |
|---|---|---|---|---|
| Solar panel | Single crystal panel | 20 | 27000000 | 540000000 |
| Inverter | 5 kW | 1 | 160000000 | 160000000 |
| Brackets | Iron corner | 20 | 1500000 | 30000000 |
| Solar tracker | In two directions | 20 | 10000000 | 20000000 |
| variable | Wire | | 20000000 | 20000000 |
| total sum | Mobile two-way system | | | 770000000 |

**Table 2: System Requirements (Computed and Optimized by PVsyst)**

## Simulation and its economic study in fixed panel mode:

Using the geographical information of the power plant installation site, in PVsyst software, the result is the best angle of placement of the panels in a fixed state, 20 degrees with an azimuth angle of 0 degrees (facing south), which in all simulations of this state has used this angle to obtain the highest power.

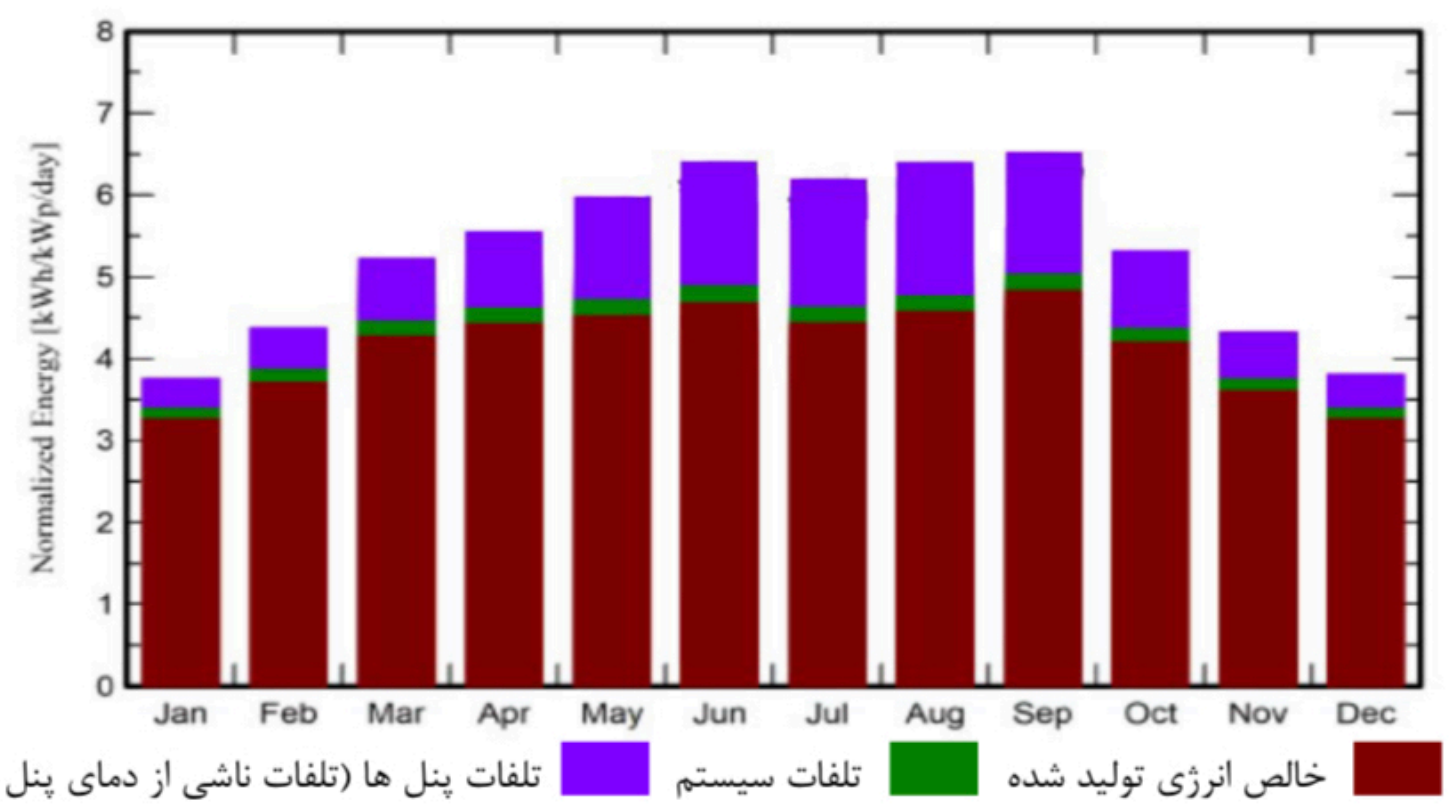

**Figure 6: Average daily energy produced in each month of the year**

The production capacity of the system in this case is 10,030 kWh, taking into account the losses from the system and the temperature of the panels, the net energy delivered can be distributed to the national electricity grid is 8,060 kWh per year. The occupied area in this case is 60 meters according to standards. It is necessary to install the panels to prevent them from casting shadows on top of each other. Heat losses in the panels section have increased a lot in recent months. According to software calculations, these losses have been produced at an average of 13% of the total energy throughout the year. Losses in the system, wiring and inverter are shown in green in Figure 6, by placing them in sections away from direct sunlight and applying the standards mentioned in the above sections, we will have less losses. It is not possible to place the panels away from direct solar radiation to prevent the temperature from rising. Solar radiation is directly related to the increase in electricity production in the system. They do not produce gas or pollution, using this system will prevent the production of only 82.6 tons of 2 co gas and other greenhouse gases in one year.

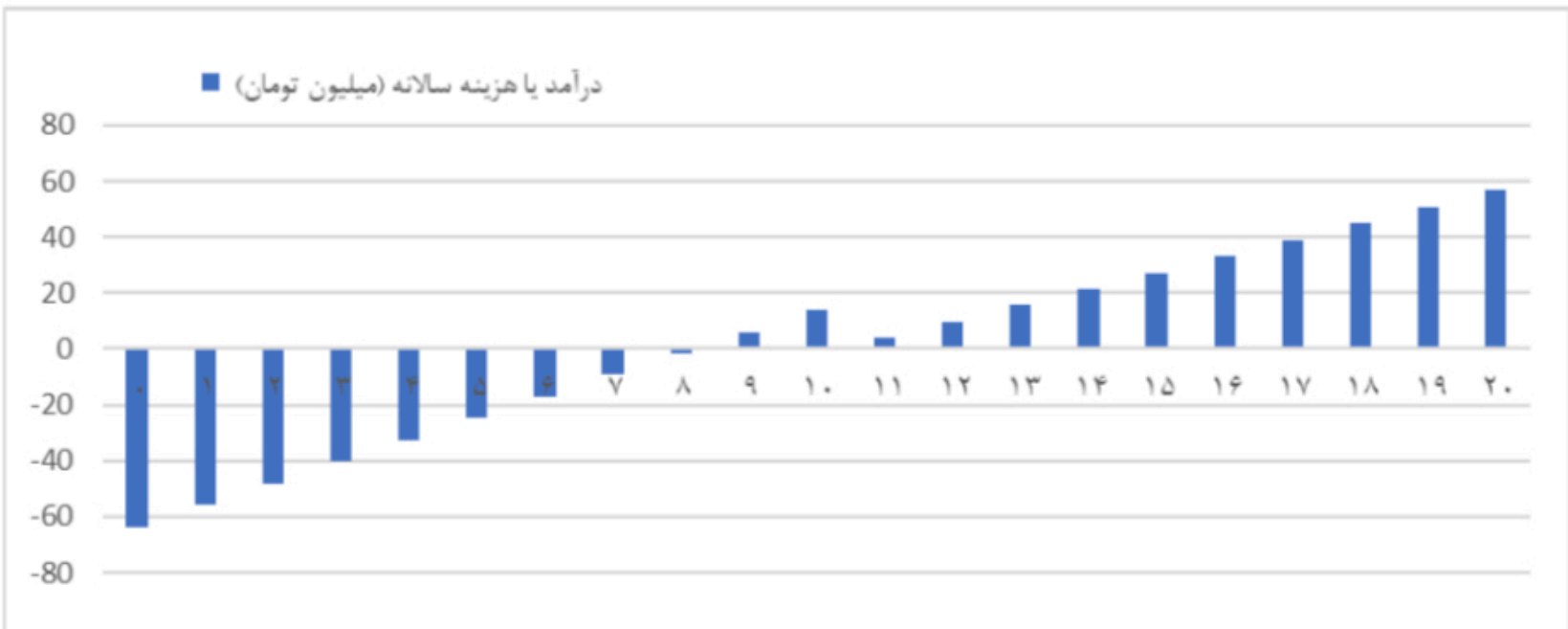

**Figure 7: Current value of the project**

From the current value chart, we will have Figure 7 of the project, which will be profitable from the year 9 onwards. The reason for the decrease in the chart in 10 to 12 years is the Cost of inverter replacement. Simulation and economic study in the case of a system equipped with a solar tracker and moving panels in two directions:

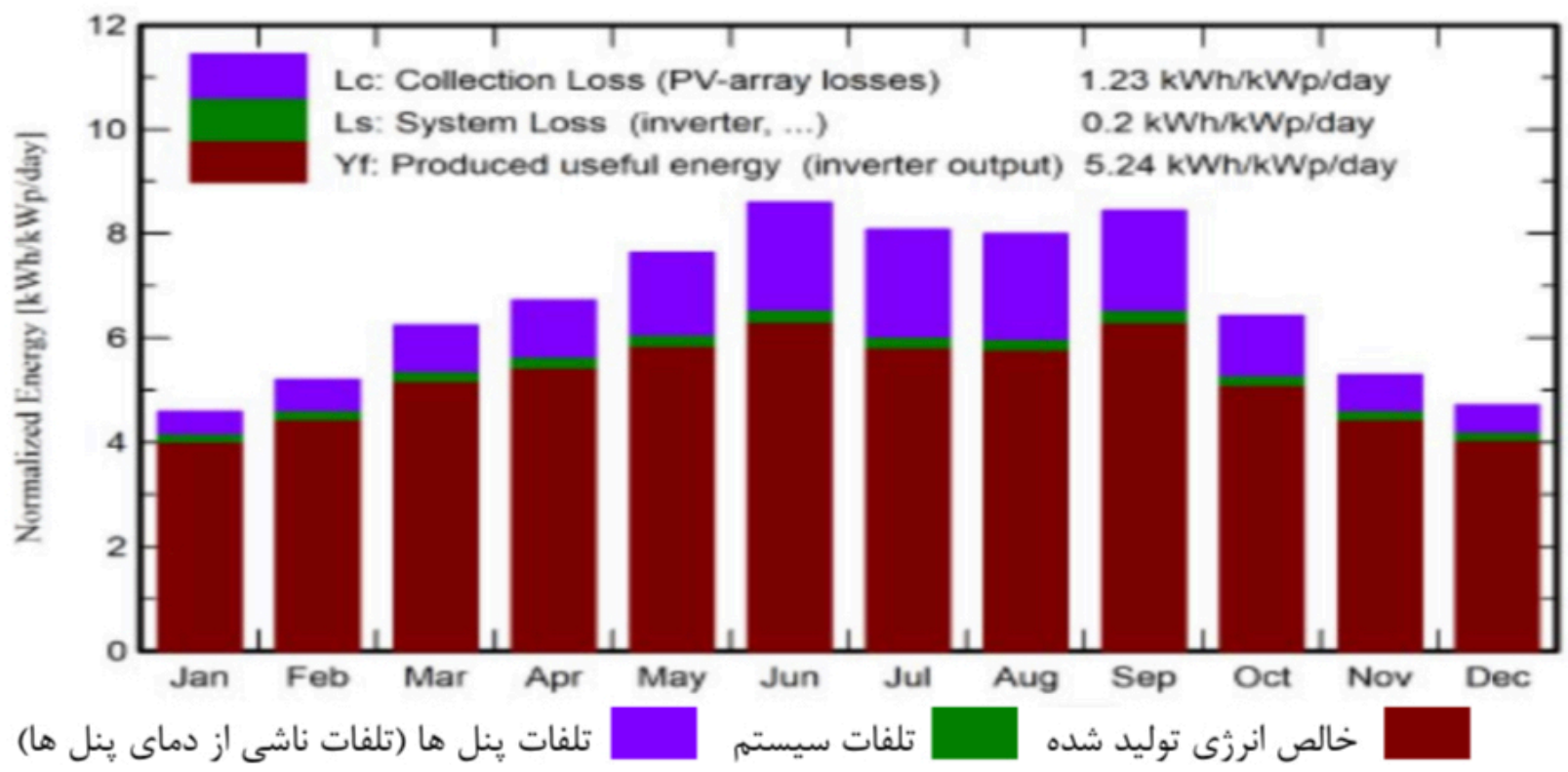

**Figure 8: Average daily energy produced in each month of the year**

In this case, the total power generated without taking into account the losses of 12,750 kWh per year, in terms of system losses and the effect of temperature losses of panels to 10,100 kWh per year, net production can be distributed to the national grid. The occupied area in this case is 80 meters; the increase in space is due to the mobility of the panels. As in the previous case, most of the losses belong to solar panels, with a more direct exposure to sunlight and increasing the temperature of the panels, 14.5% of the average losses of panels per year has been calculated by the software, losses due to increasing panel temperature Due to more direct exposure to sunlight, it has increased by 1%, which shows the direct relationship between solar radiation and increasing the efficiency of the panels, but due to the increase in temperature, temperature losses also increase. In this case, the production of 106.4 tons 2 CO and other greenhouse gases are prevented.

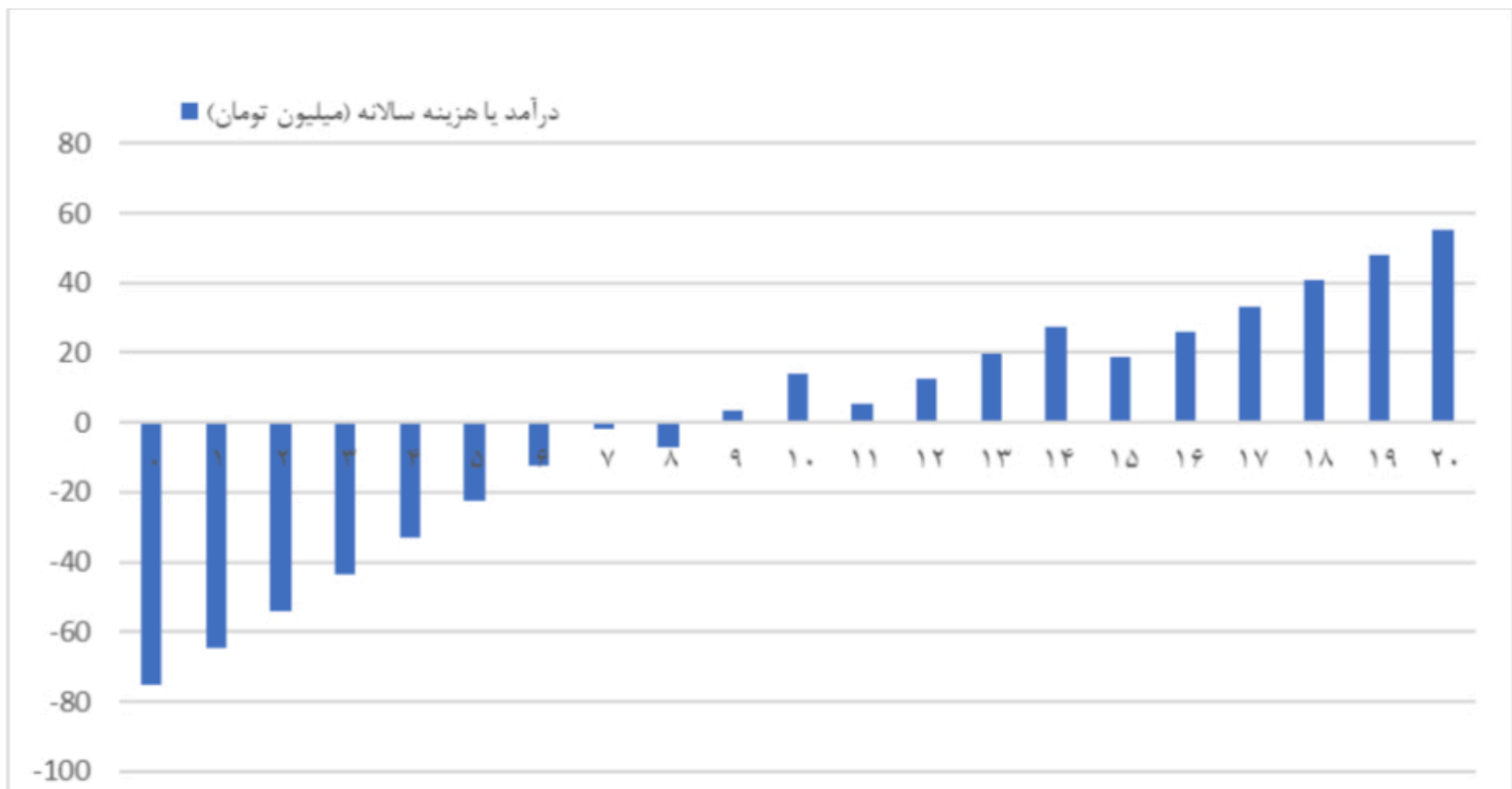

**Figure 9: Current value of the project**

In this case, the project will be profitable according to Figure 4-2-2 from 9 onwards, the reason for the decline in value flow. The current in the ranges 6 to 8 and 14 to 16 is related to the cost of replacing solar trackers, in the range 10 to 12 is related to the replacementIt is an inverter. The rate of return in this case is 173% in the next twenty years.

## Conclusion:

Purpose in designing photovoltaic systems, by each of the common patterns, designing photovoltaic systems with it has the highest efficiency. Different conditions are influential in achieving this goal, weather conditions, and radiation levels the sun in our dear country Iran, this amount in all parts of the country is an average of 4.5 hours per day, even from The global average is also higher. It is possible to design photovoltaic systems in almost any city. In tropical cities, with increasing temperature, we see an increase in load consumption in the urban electricity system and to reduce this amount of pressure Utilization of 1 to 5 kW photovoltaic systems is suitable as scattered micro grids, these micro grids due to Occupying little space, they can be easily installed on rooftops. The average occupied area in both methods is approximately 70 meters; it can be easily installed on the roofs. Considering the guaranteed purchase of electricity in photovoltaic systems, it is quite economical. The current value method is a more economical project for a project with a higher current value. Solving this problem is a good way to build a photovoltaic power plant on the water surface.

## Resources:

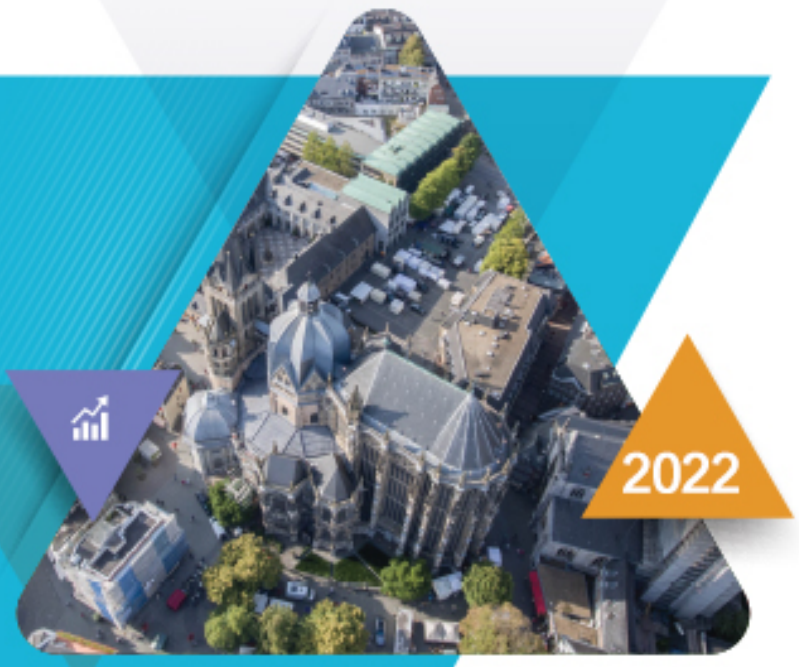